\documentclass[aps,prd,preprint,superscriptaddress,nofootinbib,showpacs]{revtex4}
\usepackage{graphicx}
\usepackage{verbatim}
\usepackage{ulem}
\usepackage{color}
\definecolor{My_red}        {cmyk}{0.00,1.00,1.00,0.20}

\usepackage{braket}

\newcommand{\bmat}{\left(\begin{array}}
\newcommand{\emat}{\end{array}\right)}
\newcommand{\beq}{\begin{equation}}
\newcommand{\eeq}{\end{equation}}

\def\ra{\rightarrow}

\def\ld{\lambda}
\def\f{\frac}

\def\bwt{\begin{widetext}}
\def\ewt{\end{widetext}}
\def\be{\begin{equation}}
\def\ee{\end{equation}}
\def\bea{\begin{align}}
\def\eea{\end{align}}
\def\bean{\begin{align*}}
\def\eean{\end{align*}}
\def\bary{\begin{array}}
\def\eary{\end{array}}
\def\bit{\begin{itemize}}
\def\eit{\end{itemize}}

\def\ra{\rightarrow}

\def\ld{\lambda}

\def\su5u1{SU(5) \times U(1)}
\def\fsu5u1{SU(5) \times U(1)'}
\def\so10{SO(10)}
\def\sq20{SO(10) \times SO(10)}

\def\ra{\rightarrow}

\def\ld{\lambda}
\def\f{\frac}

\def\L{\left(}
\def\R{\right)}

\def\ra{\rightarrow}

\def\ld{\lambda}

\def\su5u1{SU(5) \times U(1)}
\def\fsu5u1{SU(5) \times U(1)'}
\def\so10{SO(10)}
\def\sq20{SO(10) \times SO(10)}

\usepackage[centertags]{amsmath}
\usepackage{amssymb}

\begin{document}

\title{Light Doubly Charged Higgs Boson via the $WW^*$ Channel at LHC}

\author{Zhaofeng Kang}
\email{zhaofengkang@gmail.com}
\affiliation{School of Physics, Korea Institute for Advanced Study,
Seoul 130-722, Korea}
\affiliation{Center for High-Energy
Physics, Peking University, Beijing, 100871, P. R. China}

\author{Jinmian Li}
\email{phyljm@gmail.com}
\affiliation{ARC Centre of Excellence for Particle Physics at the Terascale, Department of Physics, University of Adelaide, Adelaide, SA 5005, Australia}
\affiliation{State Key Laboratory of
Theoretical Physics and Kavli Institute for Theoretical Physics
China (KITPC), Institute of Theoretical Physics, Chinese Academy of
Sciences, Beijing 100190, P. R. China}

\author{Tianjun Li}
\email{tli@itp.ac.cn}
\affiliation{State Key Laboratory of
Theoretical Physics and Kavli Institute for Theoretical Physics
China (KITPC), Institute of Theoretical Physics, Chinese Academy of
Sciences, Beijing 100190, P. R. China}
\affiliation{School of Physical Electronics, University of
Electronic Science and Technology of China, Chengdu 610054, P. R.
China}

\author{Yandong Liu}
\email{ydliu@itp.ac.cn}
\affiliation{State Key Laboratory of
Theoretical Physics and Kavli Institute for Theoretical Physics
China (KITPC), Institute of Theoretical Physics, Chinese Academy of
Sciences, Beijing 100190, P. R. China}

\author{Guo-Zhu Ning}
\email{ngz@mail.nankai.edu.cn}
 \affiliation{Center for High-Energy
Physics, Peking University, Beijing, 100871, P. R. China}

\date{\today}

\begin{abstract}

The doubly charged Higgs bosons $H^{\pm\pm}$ searches at the Large Hadron Collider (LHC) have been studied extensively and strong bound is available for $H^{\pm\pm}$ dominantly decaying into a pair of same-sign di-leptons. In this paper we point out that there is a large cavity in the light $H^{\pm\pm}$ mass region left unexcluded. In particular, $H^{\pm\pm}$ can dominantly decay into $WW$ or $WW^*$ (For instance, in the type-II seesaw mechanism the triplet acquires a vacuum expectation value around 1 GeV.), and then it is found that $H^{\pm\pm}$ with mass even below $2m_W$ remains untouched by the current collider searches. Searching for such a $H^{\pm\pm}$ at the LHC is the topic of this paper. We perform detailed signal and background simulation, especially including the non-prompt $t\bar{t}$ background which is the dominant one nevertheless ignored before. We show that such $H^{\pm\pm}$ should be observable at the 14 TeV LHC with 10-30 fb$^{-1}$ integrated luminosity.

\end{abstract}

\maketitle
\section{introduction}

At the large hadron collider (LHC), the searches for new physics beyond the standard model (SM) have a preference for the colored particles. It is due to two reasons. First, from the argument for solving the gauge hierarchy problem,  colored partners of top quark are expected, to cancel the quadratic divergence of Higgs mass incurred by top quark. Second, viewing from detectability, colored particles  have sizable production rates even at the well motivated TeV scale. Nevertheless, it is also of importance to investigate the status and prospects of new electroweak (EW)  particles. They are not less motivated in particle physics. But at the LHC these particles, typically with small production rates, are inclined to be buried in the huge SM EW and/or QCD backgrounds, except for those with characterized signatures, e.g., large missing transverse energy or same-sign di-lepton (SSDL). The latter frequently originates from particles with a larger electric charge, and the doubly charged Higgs bosons, denoted as $H^{\pm\pm}$, is a good case in point.

A lot of works have been done on the LHC search for $H^{\pm\pm}$ that come from the (scalar) $SU(2)_L$ triplet representation with hypercharge $\pm1$ (denoted as $\Delta$).~\footnote{$H^{\pm\pm}$ can also be arranged in a singlet~\cite{Zee:1985id}, doublet~\cite{Law:2013dya} $SU(2)_L$ and even higher dimensional~\cite{Cirelli:2005uq,Babu:2009aq,Cai:2011qr} representations. Some of them may produce similar signatures studied in this paper.} As a matter of fact, extension to the SM Higgs sector by $\Delta$ is well inspired by various new physics contexts, e.g., solving the hierarchy problem~\cite{GMM,little}, providing a viable dark matter candidate~\cite{FileviezPerez:2008bj} and in particular generating neutrino masses via the seesaw mechanism~\cite{type2}. In supersymmetry, such triplets provide an effective way to lift the SM-like Higgs boson mass, thus greatly relieving the fine-tuning problem~\cite{Kang:2013ft}. In addition, a light $\Delta$ on the loop of Higgs decay into a pair of photon may appreciably affect the corresponding branching ratio~\cite{Arhrib:2011vc,Chun:2013ft,Kang:2013ft,Dev:2013ff}; it would be of particular interest if we were at  the early stage of LHC, which hinted a sizable di-photon excess.

Most of the previous works on $H^{\pm\pm}$ searches concentrate on the heavy mass region, while in this article we will focus on the complementary region, the light mass region, i.e. lighter than $2m_W$ but above $m_W$. Extensive attentions are paid on the decay modes of $H^{\pm\pm}$ dominated by either the SSDL~\cite{Perez:2008ha,Rentala:2011mr} or di-$W$ \cite{Han:2007bk, Chiang, Ding:2014nga}, or the cascade decay among scalar fields \cite{H_cas,Chakrabarti:1998qy,Han:2015}. For a comprehensive discussion on the relative importance of the decay channels of $H^{\pm\pm}$,  see Ref.~\cite{Melfo:2011nx}. The search for $H^{\pm\pm}$ through the SSDL channel has been peformed at the LHC, which already excludes the mass of $H^{\pm\pm}$ up to about 300 GeV~\cite{Chatrchyan:2012ya,ATLAS:2012hi}. However, in the current experimental searches  other decay modes like di-$W$ may still allow a much lighter $H^{\pm\pm}$~\cite{Kanemura:2013vxa}, for instance, even below $2m_W$. Note that such $H^{\pm\pm}$ decays into di-$W$ with one being off-shell, thus this channel is dubbed $WW^*$.

Mainly owing to the softness of the final products, hunting for $H^{++}\ra WW^*$ is a challenging task at LHC even with merits of relatively large  pair production cross section and  the remarkable SSDL signature. So it is very important to elaborate the LHC search for such light $H^{\pm\pm}$. We shall perform the detailed background simulation on SSDL, especially including the non-prompt $\bar t t$ background which is the dominant one nevertheless ignored before. We find that $H^{\pm\pm}$ should be observable at the 14 TeV LHC with $10-30\rm\,fb^{-1}$ integrated luminosity. The last but not the least, here we take a simplified model approach and discuss the search for $H^{\pm\pm}$ in the simplified model at the LHC, which makes our result less model-dependent and can be conveniently translated into
other specific models~\cite{Rentala:2011mr,Alves:2011wf}.

This paper is organized as follows. In Section~\ref{model}, we describe some details about the simplified model for
the doubly charge Higgs bosons in $SU(2)_L$ triplet representation and consider some relevant constraints. Section~\ref{property} is devoted to the properties of the doubly charged Higgs bosons including its productions
and decays at the LHC. In Section~\ref{LHCsearch}, we study the detailed collider simulation for both signal
and background events,  and present the LHC reach of the doubly charged Higgs boson.
Finally we conclude and give a outlook in Section~\ref{conclusion},
 and some necessary details are given in Appendix A.

\section{The SM Extension with a Hypercharge $Y=\pm1$  Triplet Higgs} \label{model}

\subsection{The simplified model}

There are a lot of motivated new physics models which have a $SU(2)_L$ triplet Higgs boson $\Delta$ with hypercharge $Y=\pm1$. In order to make our discussion as general as possible, in this work we take the simplified model approach and make the assumption that in the simplified model new particles other than $\Delta$ are absent or decoupled. Thus, the relevant terms in the Lagrangian can be written as 
\begin{eqnarray}
  \mathcal{L}\supset\mathcal{L}_{\rm kin}+\mathcal{L}_Y-V(\Phi,\Delta),
\end{eqnarray}
where $\mathcal{L}_{\rm kin}, \mathcal{L}_Y$ and $V(\Phi,\Delta)$ are the kinetic term, the Yukawa interaction, and the Higgs potential, respectively. Let us define the SM Higgs doublet and the triplet as
\begin{eqnarray}\label{fields}
\Phi = \left( \begin{array}{c}
                            \phi^+ \\
                            \phi^{0}
                            \end{array} \right),\;\;
\Delta = \left( \begin{array}{cc}
                            \frac{\delta^+}{\sqrt{2}} & \delta^{++} \\
                            \delta^{0} &  -\frac{\delta^+}{\sqrt{2}}
                            \end{array} \right),
\end{eqnarray}
with $\phi^0=\frac{1}{\sqrt{2}}(\phi+v_\phi+i\chi),\; \delta^0=\frac{1}{\sqrt{2}}(\delta+v_\Delta+i\eta)$. 

Generically, the scalar potential $V(\Phi,\Delta)$ generates a non-vanishing vacuum expectation value (VEV) $v_\Delta$ for the neutral component of $\Delta$. The most general scalar potential is
\begin{eqnarray}\label{higgspotential}
V(\Phi,\Delta) &=& m^2 \Phi^{\dagger}\Phi+ M^2 {\rm Tr}
(\Delta^{\dagger}\Delta) + \lambda_1 (\Phi^{\dagger}\Phi)^2 + \lambda_2 [{\rm Tr}(\Delta^{\dagger}\Delta)]^2 + \lambda_3 {\rm Tr}[(\Delta^{\dagger}\Delta)^2]  \nonumber \\
&& + \lambda_4 (\Phi^{\dagger}\Phi){\rm Tr}(\Delta^{\dagger}\Delta) +
\lambda_5 \Phi^{\dagger}\Delta \Delta^{\dagger} \Phi
 + \left[\mu (\Phi^{\intercal} \mathrm{i}\tau_2 \Delta^{\dagger}\Phi) + h.c.\right].
\end{eqnarray}
If $\mu=0$, the potential will respect a $Z_2$ symmetry acting on $\Delta$ and the triplet may do not acquire VEV. Otherwise, $\delta^0$ is supposed get a non-vanishing VEV. After minimizing the potential Eq. (\ref{higgspotential}) and considering very small $v_\Delta$ (grounded on reason discussed soon later), one gets 
\begin{eqnarray}
v_{\Delta} \simeq \f{\mu}{\sqrt{2}} \frac {v^2_\phi}{ M^2+\frac{1}{2}\L\ld_4+\ld_5\R v^2_\phi}
=\f{\mu}{\sqrt{2}} \frac {v^2_\phi}{ M^2_\Delta}.
\end{eqnarray}
We can see that there are typically two ways to achieve a sufficiently small $v_{\Delta}$: (A) $\mu$ is around the weak scale, and then the triplet is pushed up to the TeV region; (B) by contrast, the triplet is around the weak scale with $M_\Delta\sim v_\phi=246$ GeV, and then $\mu$ is forced to lie below the GeV scale as $\mu=v_\Delta$.\footnote{Since as $\mu\ra0$ a symmetry arises, this case is at least technically natural according to the 't Hooft principle.} 


We now explain why $v_\Delta$ is restricted to be very small. The Higgs kinetic terms are
\begin{eqnarray}\label{kinetic1}
  \mathcal{L}_{\rm kin}\supset(D_\mu\Phi)^\dagger(D^\mu\Phi)+{\rm Tr}\left[D_\mu\Delta)^\dagger(D^\mu\Delta)\right],
\end{eqnarray}
where the covariant derivatives are defined by
\begin{eqnarray}\label{kinetic2}
  D_\mu\Phi=\left(\partial_\mu+i\frac{g}{2}\tau^aW^a_\mu+i\frac{g^\prime}{2}B_\mu\right)\Phi,
  \;\;
  D_\mu\Delta=\partial_\mu\Delta+i\frac{g}{2}[\tau^aW^a_\mu,\Delta]+ig^\prime B_\mu\Delta,
\end{eqnarray}
with $(W^a_\mu,g)$ and $(B_\mu, g^\prime)$ are, respectively, the $SU(2)_L$ and $U(1)_Y$ gauge fields and couplings, and $\tau^a=\sigma^a/2$ with $\sigma^a (a=1,2,3)$ the Pauli matrices. According to Eqs.~(\ref{fields}), (\ref{kinetic1}) and (\ref{kinetic2}), the masses of the $W$ and $Z$ gauge boson at tree level are
\begin{eqnarray}
  m^2_W=\frac{g^2}{4}(v^2_\phi+2v^2_\Delta),\;\; m^2_Z=\frac{g^2}{4\cos\theta_W}(v^2_\phi+4v^2_\Delta).
\end{eqnarray}
Asides from the SM contributions, they receive additional contributions from the triplet. As a consequence, the oblique parameter $\rho$ will be modified. Now, it is given by
\begin{eqnarray}
  \rho=\frac{m^2_W}{m^2_Z\cos^2\theta_W}=\frac{1+2x^2}{1+4x^2}\approx1-2x^2,
\end{eqnarray}
with $x=v_\Delta/v_\phi$. The current experimental value of $\rho$~\cite{Beringer:1900zz} imposes a strict constraint on the deviation of $\rho$ from 1 and yields the upper bound $x\lesssim 0.01$, or in other words, $v_\Delta\lesssim 2.46$ GeV. We will turn back to this latter.

Although almost irrelevant to our later LHC studies, we for completeness still incorporate the Yukawa interactions of the triplet field, which are crucial in generating neutrino masses in type-II seesaw mechanism.~\footnote{In this paper we will use this model as the benchmark model for the completion of the simplified model.} It takes the form of
\begin{eqnarray} \label{yukawa}
-\mathcal{L}_Y  &\supset& y_{ij} L^{T}_{i} \mathcal{C} i \tau_2 \Delta L_{j} + h.c.\nonumber \\
&=&y_{ij}\left[\nu^T_i \mathcal{C}P_L\nu_j\delta^0- \frac{1}{\sqrt{2}}(\nu^T_i \mathcal{C}P_L\ell_j-\ell^T_i \mathcal{C}P_L\nu_i)\delta^+ -\overline{\ell^C_i}P_L\ell_j\delta^{++}\right]
+h.c.~,~\,
\end{eqnarray}
where $y_{ij}(i,j=1,\,2,\,3)$ is an arbitrary symmetric complex matrix, $\mathcal{C}=i\gamma^0\gamma^2$ is the charge conjugation operator, and $L^T_i=(\nu_{iL},\ell_{iL})$ is a left-handed lepton doublet in the SM. After the EW symmetry breaking, the Majorana neutrino mass terms are generated
\begin{eqnarray*}
  (M_\nu)_{ij}=\sqrt{2} y_{ij} v_\Delta~.
\end{eqnarray*}


To end up this subsection, we give a quick recapitulation of the scalar mass spectrum. In addition to the three Nambu-Goldstone  $G^\pm$ and $G^0$ which are absorbed by the longitudinal components of the $W^\pm$ and $Z$ gauge bosons, the model has seven physical Higgs bosons ($H^{\pm\pm}, H^{\pm}, H^0, A^0$, and $h$). The doubly charged Higgs $H^{\pm\pm}$ is purely from the triplet ($H^{\pm\pm}=\Delta^{\pm\pm}$), while the other Higgs bosons would be in general mixtures of the SM Higgs and triplet fields. Such mixings are proportional to $x$ and hence seriously suppressed. For simplicity, the masses of these triplet-like Higgs bosons are collected together as
follows (neglecting $\mathcal{O}(v^2_\Delta/v^2_\phi)$ terms)
\begin{eqnarray}\label{spectum}
M_{H^{\pm \pm}}^2 &\approx& M^2_\Delta - \frac{1}{2}\lambda_5v^2_\phi~,~
\nonumber\\
M_{H^{\pm}}^2 &\approx &M^2_\Delta - \frac{1}{4}\lambda_5v^2_\phi ~,~ \nonumber\\
M_{H,A}^2 &\approx & M^2_\Delta~.~
\end{eqnarray}
So we can see that the quartic $\lambda_5-$term is responsible for the masses splittings, which satisfy the relations
\begin{eqnarray}\label{mass}
  M_{H^{\pm \pm}}^2-M_{H^{\pm}}^2=M_{H^{\pm}}^2-M_{H,A}^2=-\frac{1}{4}\lambda_5v^2_\phi.
\end{eqnarray}
It is shown that there exits three patterns of the mass spectrum for the triplet-like Higgs bosons. When $\lambda_5=0$, all the triplet-like Higgs bosons are degenerate in mass. However, in the case $\lambda_5>0$ ($\lambda_5<0$), the resulting mass orderings become $M_{H,A}>M_{H^\pm}>M_{H^{\pm\pm}}$ ($M_{H,A}<M_{H^\pm}<M_{H^{\pm\pm}}$).

\subsection{Possible constraints}

There are various possible theoretical and experimental constraints on the triplet Higgs model or Type-II seesaw
model~\cite{Arhrib:2011uy,Chun:2012jw,Aoki:2012jj,Queiroz:2014zfa}. Here, we only include some constraints
which are closely relevant to our study.

\subsubsection{On the magnitude of $v_\Delta$}

As discussed above, the VEV $v_\Delta\neq 0$ modifies the tree-level relation for the electroweak $\rho$ parameter as $\rho\approx 1-2 v^2_\phi/v^2_\Delta $. However, this mass splittings between the component of $\Delta$ will induce an additional positive contribution, with proportional to mass splitting, to $\rho$ to cancel the effect lead by $v_\Delta$, for example, an upper limit from perturbativity ($\lambda_5\lesssim 3$) to be $v_\Delta\lesssim 7 ~{\rm GeV}$, for $m_H=120~{\rm GeV}$ \cite{Melfo:2011nx}. Conservatively, we take the upper bound  $v_\Delta \lesssim 2 ~{\rm GeV}$, which is corresponding to $x=v^2_\phi/v^2_\Delta\lesssim 0.01$.

The lepton flavor violations involving $\mu$ and $\tau$ provide the strongest constraint on the $y_{ij}$ and thus $v_\Delta \sim (M_\nu)_{ij}/y_{ij}$. To accommodate the currently favored experimental constraints, there is a lower limit $ v_\Delta M_{H^{\pm\pm}}\gtrsim 100 ~ {\rm eV\, GeV}$~\cite{Akeroyd:2009nu}, which is quite loose. A relevant constraint comes from the neutrino masses. If the Yukawa coupling of triplet scalar is the unique origin for neutrino mass, the current observations from the neutrino oscillation experiments and cosmological bounds give \cite{Beringer:1900zz}:
\begin{eqnarray}\label{neutrinomass}
  m_\nu=\sqrt{2}y_{ij} v_\Delta\lesssim 10^{-10} ~{\rm GeV}~.
\end{eqnarray}
For our purpose, a larger $v_\Delta$ is of interest. Then, for $v_\Delta=1\, {\rm GeV}$ one needs an extremely small $y_{ij}\lesssim 10^{-10}$ to accommodate the correct neutrino mass scales. But it is not of concern in the simplified model which is not a model for neutrino physics. For example, beyond the simplified model maybe there are some other source for generating neutrino masses and then the Yukawa couplings can be forbidden absolutely. In summary, $v_\Delta$ can be as large as 1 GeV without spoiling any constraints; moreover, the Yukawa couplings $y_{ij}$ can be made arbitrarily small in order to suppress the direct decay into a pair of lepton.

\subsubsection{Experimental bounds on $M_H^{\pm\pm}$}

The mass of doubly charged Higgs $M_H^{\pm\pm}$ has been constrained in the past experiments such as SLC and LEP, independently of the decay modes of $H^{\pm\pm}$. From the LEP experiment, the width of $Z$ boson has been precisely measured. When $M_H^{\pm\pm}$ is less than  half of the $Z$ boson mass, the new decay mode $Z \rightarrow H^{\pm\pm}H^{\mp\mp}$ will open. Then the total decay width of the $Z$ boson will receive a sizable contribution from the partial width as
\begin{eqnarray}
\Gamma(Z \rightarrow H^{\pm\pm}H^{\mp\mp}) = \frac{G_Fm_Z^3}{6\pi\sqrt{2}}(1-2s_W^2)^2\L1-\frac{4M_{H^{\pm\pm}}^2}{m_Z^2}\R^{\frac{3}{2}}~.~
\end{eqnarray}
On the other hand, from \cite{Beringer:1900zz} we know
\begin{eqnarray}
\Gamma_Z^{NP} < 3 ~ \textrm{MeV}~ (95\% \textrm{CL})~,
\end{eqnarray}
and this puts a stringent constraint on the mass of doubly charged scalar. The lower mass bound can be obtained $M_{H^{\pm\pm}} > 42.9$ GeV at $95\%$ confidential level.

The mass bound on $M_H^{\pm\pm}$ can also be taken through its direct searches at the LHC.  The ATLAS Collaboration has searched for doubly-charged Higgs bosons via pair production in the SSDL channel. Based on the data sample corresponding to an integrated luminosity of 4.7 $\text{fb}^{-1}$ at $\sqrt{s} = 7$ TeV, the masses below 409 GeV, 375 GeV and 398 GeV have been excluded respectively for $e^{\pm}e^{\pm}$, $e^{\pm}\mu^{\pm}$ and $\mu^{\pm}\mu^{\pm}$ by assuming a branching ratio of $100\%$ for each final state \cite{ATLAS:2012hi}. Besides pair production, the CMS Collaboration also considered the associated production $pp \rightarrow H^{\pm\pm}H^{\mp}$, in which the masses of $H^{\pm\pm}$ and $H^{\mp}$ are assumed to be degenerate. Using three or more isolated charged lepton final states, the upper limit on $M_{H^{\pm\pm}}$ is driven under specific assumptions on branching ratios \cite{Chatrchyan:2012ya}. However, other decay modes for $H^{\pm\pm}$ such as di-$W$ will become dominant under some conditions. The preliminary search for doubly-charged Higgs boson based on this channel is also studied in Ref.~\cite{Kanemura:2013vxa}. By fully utilizing the result of the SSDL search by the ATLAS Collaboration (with $4.7~ \text{fb}^{-1}$ integrated luminosity at $\sqrt{s} = 7$ TeV), the lower limit is obtained to be 60 GeV at the $95\%$ C.L.. Moreover, considering the integrated luminosity of $20~ \text{fb}^{-1}$, the lower bound can be evaluated to 85 GeV. Since the treatment for backgrounds and signals in the $WW^*$ channel will be in principle different from the SSDL case, a detailed analysis on this topic is necessary. In this article, we concentrate on this scenario and elaborate the search for such a $H^{\pm\pm}$ at LHC.

\section{Production and decay of $H^{\pm\pm}$}\label{property}

\subsection{Production}

The prospect for  the production of doubly charged scalar $H^{\pm\pm}$ has been widely studied at the hadron colliders such as Tevatron and LHC. For an elaborate discussion on this topic, please see \cite{Rentala:2011mr}. 
The main production processes for $H^{\pm\pm}$ at the LHC are the pair production via Drell-Yan process $p p \rightarrow \gamma^\ast /Z \rightarrow H^{\pm\pm} H^{\mp\mp}$  and the associated production $p p \rightarrow W^{\pm\ast} \rightarrow H^{\pm\pm} H^{\mp}$. Note that these processes only depend on the mass of the doubly charged Higgs boson $m_{H^{\pm\pm}}$ and independent on $v_\Delta$; even it is as large as 1 GeV.  The next-to-leading (NLO) QCD corrections to the pair production can increase the cross section by about $20-30\%$ \cite{Muhlleitner:2003me}. Moreover, the authors have calculated the two-photon fusion process and found its contribution to the pair production can be comparable with the NLO QCD corrections to the Drell-Yan process \cite{Han:2007bk}. For conservatively, we only consider the leading-order (LO) cross section in this work.
\begin{figure}[htb]
 \begin{center}
 \includegraphics[scale=0.5]{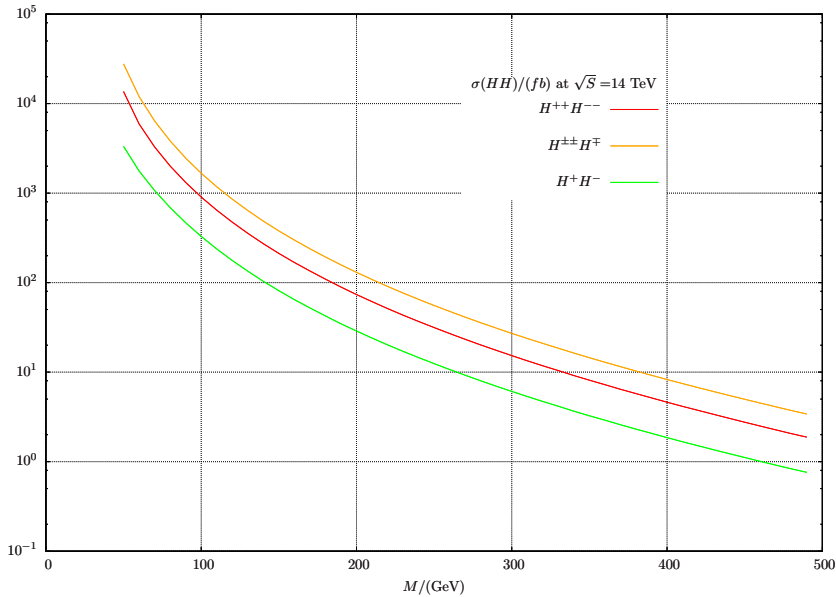}
 \caption{\label{xsec} The leading order production cross sections for $H^{\pm\pm}H^{\mp\mp}$, $H^{\pm\pm}H^{\mp}$ and $H^{\pm}H^{\mp}$ at
the 14 TeV LHC. We assume the degenerate mass of $H^{\pm\pm}$ and $H^{\pm}$ for $H^{\pm\pm}H^{\mp}$ associate production. }
 \end{center}
 \end{figure}

In Fig~\ref{xsec}, we show the LO production cross sections for the corresponding charged Higgs pair productions at the 14 TeV LHC. The production rate ranges from a few fbs to a few pbs in the mass range of [50, 500]  GeV.
We have also shown in this figure the production rate of $H^{\pm\pm}H^{\mp}$ associated production, assuming mass degeneracy between $H^{\pm\pm}$ and $H^\pm$, whose rate is a few times larger than the $H^{\pm\pm}H^{\mp\mp}$ pair production. Hereafter, we only consider the $H^{\pm\pm}H^{\mp\mp}$ pair production as a more conservative study.

\subsection{Decays}

In the simplified model given in the previous section, the possible decay modes for a light $H^{\pm\pm}$ considered in this paper include: (1) the lepton-number violating (LNV) decay mode $H^{\pm\pm}\rightarrow \ell^\pm_i\ell^\pm_j$; (2) the $WW^*$ decay mode $H^{\pm\pm}\rightarrow W^\pm W^{\pm\ast} \rightarrow W^\pm f\bar{f} $; (3) the cascade decay mode $H^{\pm\pm}\rightarrow H^\pm W^{\pm\ast} \rightarrow W^\pm f\bar{f}$. The corresponding decay rates can be found in Appendix \ref{DHdecay}. In particular, in the models with type-II seesaw mechanism, the LNV decays are proportional to Yukawa coupling $y_{ij}$, consequently inversely proportional to $v_\Delta$ due to $v_\Delta=M_v/y$. In contrast, the $WW^*$ mode is proportional to $v_\Delta$, which means that the higher the value of $v_\Delta$ is, the more important we expect the $WW^*$ mode to be, with a corresponding decrease in the LNV. As for the cascade decay mode, it is induced by the gauge interactions and highly sensitive to the mass splitting $\Delta M= M_{H^{\pm\pm}}-M_{H^{\pm}}$.

\begin{figure}[htb]
 \begin{center}
 \includegraphics[scale=1.0]{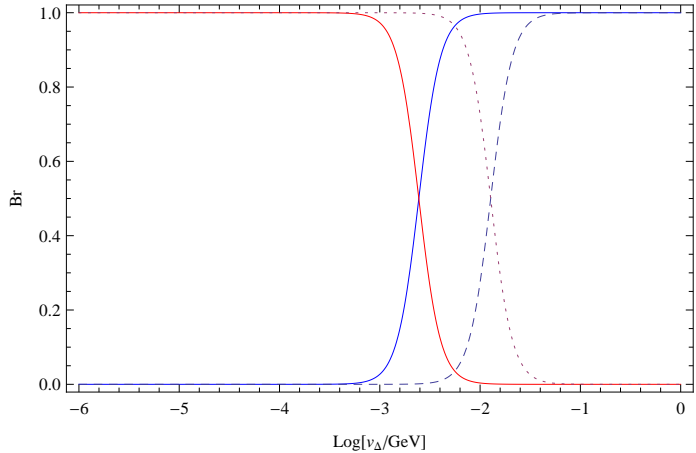}
 \caption{\label{BRvd} The branching ratios of the doubly charged Higgs boson decay versus $v_\Delta$ for $M_{H^{\pm\pm}}=100 {\rm GeV}$ (dash line) and $M_{H^{\pm\pm}}=150 {\rm GeV}$ (solid line). The red and blue lines are for the LNV decays and $WW^*$ mode,
respectively. }
 \end{center}
 \end{figure}

\begin{figure}[htb]
 \begin{center}
 \includegraphics[scale=1.0]{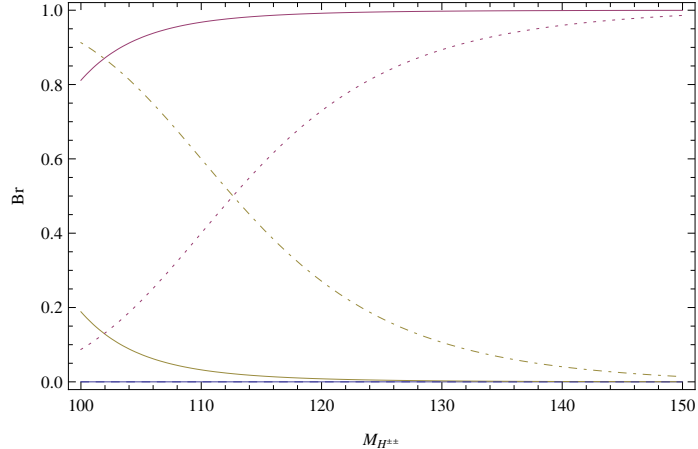}
 \caption{\label{BRmdh} The branching ratios of the doubly charged Higgs boson decay versus $M_{H^{\pm\pm}}$ for $\Delta M=2 {\rm GeV}$ (solid line) and $\Delta M=5 {\rm GeV}$ (dash line) with $v_\Delta=1 {\rm GeV}$. The yellow, red, and blue
lines are for the cascade decays, di-W mode, and LNV decays, respectively. }
 \end{center}
 \end{figure}

To be quantitative, the mentioned facts above have been demonstrated in Fig.~\ref{BRvd} and Fig.~\ref{BRmdh}. In Fig.~\ref{BRvd}, it is shown that, with the degenerate mass spectrum of triplet-like Higgs bosons, a relatively large $v_\Delta$ with $v_\Delta=1{\rm GeV}$ will lead the $WW^*$ mode to be the dominant decay channel of $H^{\pm\pm}$, when $M_{H^{\pm\pm}}$ is in the mass range of [100, 150] GeV. But degeneracy will be lifted for a sizable $\lambda_5$; see Eq.~(\ref{spectum}). Furthermore, for $\lambda_5<0$, which means $\Delta M= M_{H^{\pm\pm}}-M_{H^{\pm}}>0$, the cascade decays of $H^{\pm\pm}$ will open. We show all the possible decay modes of $H^{\pm\pm}$ in Fig.~\ref{BRmdh}. It is found that, for a relatively light $H^{\pm\pm}$, a mass splitting $\Delta M= 5{\rm GeV}$ makes the cascade decays rapidly overcome the $WW^*$ mode and become the dominant channel. Again, in the type-II seesaw, due to a relatively large $v_\Delta$ chosen here, the branching ratios for the LNV decays of $H^{\pm\pm}$ are always vanishingly small.

It is the right place to comment about the associated production $H^{\pm\pm}H^{\mp}$ with $H^\pm$ subsequently decaying into $H^{\pm\pm}$. The distribution of $H^{\pm\pm}$ can be similar with the direct $H^{\pm\pm} H^{\mp\mp}$ pair production as long as the mass splitting $\Delta M$ keeps small. What's more, as we can see from Fig~\ref{xsec}, the cross section of associate production is about 2 times larger than the pair production. Thus, when $\Delta M$ is small, the extra contribution from the associated production will possibly help the discovery of $H^{\pm\pm}$ (But still safe from the current LHC constraints which will be mentioned latter). Even though we only consider the direct $H^{\pm\pm}H^{\mp\mp}$ pair production in the following discussion, in technical view, our result can be generalized to include the associated production by rescaling.


\section{The LHC prospect of light $H^{\pm\pm}$}\label{LHCsearch}

In this Section, we first collect the current LHC searches for $H^{\pm\pm}$ using the SSDL signature and find that the light region of $H^{\pm\pm}$ in our scenario has not been probed yet. Then we conduct a detailed study of the discovery prospect for light $H^{\pm\pm}$ at the future LHC. It is found that the 14 TeV LHC is able to cover all the mass region of light $H^{\pm\pm}$, using the SSDL signature, aided by multi-jets and missing energy.

\subsection{The status of $H^{++}$ facing the SSDL searches}

The searches of $H^{\pm\pm}$ from the ATLAS and CMS Collaborations are both based on its LNV decays.    However, when $v_\Delta$ is significantly large and the mass spectrum of triplet-like Higgs bosons are nearly degenerate, $H^{\pm\pm}$ mainly decays into $WW^*$. The search for a light $H^{\pm\pm}$ via the $WW^*$ channel at LHC, using the SSDL signature, is our main aim in this work.

\begin{itemize}
\item Searching for $H^{\pm\pm}$ through the SSDL signature has been done before~\cite{ATLAS:2013tma,ATLAS:2012sna,Chatrchyan:2012ira,ATLAS:2012ai, CMS:2012xza}, and very strong bounds on $M_{H^{\pm\pm}}$ were derived. However, in those searches, besides the existence of SSDL, they required either a number of $b$-tagged jets, very large missing transverse energy $E^{miss}_T$ or very large $H_T = \sum_i p_T(j_i)+E^{\rm miss}_T$, which is the scalar sum of transverse momentum of jets and $E^{miss}_T$. However, here the light $H^{++}$ decay produces neither bottom quarks nor large  $E^{miss}_T$/ $H_T$, so those bounds can be evaded easily. The latter fact can also be seen from the top panels of Fig~\ref{distribution}. In the mass range we have considered, we have $E^{miss}_T \lesssim 100$ GeV and $H_T \lesssim 400$ GeV. 
\item Strong bounds ($\sim 400$ GeV) have also been derived for $M_{H^{\pm\pm}}$ if $H^{\pm\pm}$  directly decays into SSDL~\cite{ATLAS:2012hi,Chatrchyan:2012ya}. But in our scenario the SSDL signature comes from the consequent decay products along the $WW^*$ chain, and hence the invariant mass $m_{ll}$, which is peaked around the mass of $H^{\pm\pm}$ thus being a very efficient cut for $H^{\pm\pm} \to l^{\pm}_il^{\pm}_j$, no longer works well here;  see the panel in the middle left of Fig~\ref{distribution}. In addition to that, the rate of SSDL in our scenario is suppressed by the $W$ boson decay branch ratio. Therefore, there is no bound from these searches as well.

\item Until recently, the CMS Collaboration has searched for the SSDL signals with jets in low $E^{miss}_T$ and low $H_T$ region both with and without $b$-tagging~\cite{Chatrchyan:2013fea}. First, from the CMS data, we estimate the upper limit of new physics events in each signal region (SR), $N_i^{max}$. Then, following the similar procedure as in~\cite{Guo:2013asa}, we recast the analysis in~\cite{Chatrchyan:2013fea} and calculate our signal events in each SR, $N_i^{new}$. Finally, we denote the ratio $R\equiv \max_i\{N_i^{max}/N_i^{new}\}$, which indicates the CMS search sensitive to our signal process at the 8 TeV LHC. In other words, if our model was excluded, the cross section would be $R$ times larger than the prediction in the model. In the first row of Table~\ref{sig}, we list the value of $R$ for each $M_{H^{\pm\pm}}$. It is seen that $R\sim 4$, i.e., the production rates should be 4 times larger for discovery. Thereby, the benchmark points are free from this constraint even if we take into account the contribution from the associated production. 
\end{itemize}



\subsection{Backgrounds}

The backgrounds of the SSDL signature can be divided into three categories: real SSDL from rare SM processes, non-prompt lepton backgrounds, and opposite-sign dilepton events with charge misidentifications.
The non-prompt lepton backgrounds, which are the dominant background for SSDL, arise from events either with jets misidentifying as leptons or with leptons resulting from heavy flavor quark decay (HF fake). To suppress the non-prompt lepton backgrounds caused by jet misidentification, in our simulation we require the leptons in the final state to be both ``tight"~\cite{Aad:2011mk} and isolated, where the isolated lepton final state means that the scalar sum the transverse momentum of calorimeter energy within a cone of $R=0.3$ around the lepton excluding the lepton itself must be less than $16\%$ of lepton's $p_T$. { We find that the rate of jets mis-identified as leptons after the above requirements is highly suppressed, smaller than $\mathcal{O}(10^{-6})$.} Thus in the following analysis we only need to consider the non-prompt background from the heavy flavor quark decay, concretely, the semi-leptonic $t\bar{t}$ events with a non-prompt lepton from $b$-quark decay. {With our detector setup, the probability of an isolated lepton produced from $b$ quark decay is $\sim \mathcal{O}(0.1\%)$. }The dominant processes that genetate the SSDL in SM and their production cross sections at the 14 TeV LHC are listed in Table~\ref{smxsec}. The NLO production cross sections, except for $t\bar{t}Z$ and $W^\pm W^\pm jj$ are calculated by MCFM-6.6~\cite{Campbell:2011bn,Campbell:2012dh}. The NLO cross section of $t\bar{t}Z$ is taken from Ref.~\cite{Lazopoulos:2008de,Hirschi:2011pa,Garzelli:2011is,Kardos:2011na,Garzelli:2012bn}. 
As for $W^\pm W^\pm jj$, a conservatively estimated constant $K$-factor 1.5 is multiplied on its LO cross section which is calculated by MadGraph5~\cite{Alwall:2011uj}.

\begin{table}[htb]
\begin{center}
\begin{tabular}{|c|c|} \hline
Processes & $\sigma /$pb \\ \hline
$t\bar{t}$ & 843.338 \\ \hline
$W^+Z$ & 29.82 \\ \hline
$W^-Z$ & 18.33 \\ \hline
$ZZ$ & 16.12 \\ \hline
$W^+t\bar{t}$ & 0.507 \\ \hline
$W^-t\bar{t}$ & 0.262 \\ \hline
$Zt\bar{t}$ & $1.09$  \\ \hline
$W^+W^+ j j$ & $0.2377 \times 1.5$ \\ \hline
$W^-W^- j j$ & $0.1037 \times 1.5$ \\ \hline
\end{tabular}
\end{center}
\caption{Production cross sections of background processes at the 14 TeV LHC}
\label{smxsec}
\end{table}

Let us comment on the other subdominant backgrounds. The first is about the real SSDL from the rare SM processes. The relevant SM backgrounds involving Higgs boson are $tth$ (0.6 pb), $Wh$ (1.5 pb) and $Zh$ (0.8 pb), where the Higgs boson decays into $WW^*$ and $ZZ^*$ with branching ratio 21\% and 2.5\%, respectively. Among these, the most important background is $W_l(h\ra W_lW_j)$. Its production rate is similar with $W^\pm W^\pm j j$, whose contribution to our signal region is found to be small. The cross sections of $tt(h\ra V_lV_j)$, $W_l(h\ra Z_lZ_j)$ and $Z_l(h\ra W_lW_j)$ are at least one order of magnitude smaller than the corresponding backgrounds with similar final states which have been incorporate in our work, e.g., $ttV$ and $WZ$. Therefore,  these backgrounds can be neglected. The second is about the background due to charge mis-identification, which is dominated by the Drell-Yan processes, leptonic decay of $t\bar{t}$ and $W^+W^-$, in which the electrons undergone hard bremsstrahlung with subsequent photon conversion. As pointed out in~\cite{sslbgrate}, this kind of background usually contributes less than 5\% of the total backgrounds and thus will be neglected also.~\footnote{This background can also be suppressed by the isolated lepton requirement.}

\subsection{Event generation and analysis}

The signals and backgrounds are generated by MadGraph$5\_$v$1\_5\_11$~\cite{Alwall:2011uj}, where Pythia6~\cite{Sjostrand:2006za} and Delphes$\_3.0.9$~\cite{deFavereau:2013fsa} have been packed to implement parton shower and detector simulation. We implement the simplified model for doubly charged Higgs in FeynRules~\cite{Alloul:2013bka}, generating the UFO format of this model for MadGraph. Some important details in our simulation are summarized here. 
In the first, the matrix element of signals and all backgrounds, except for $W^\pm W^\pm j j$, are generated up to 2 jets. 
Next, we use the MLM matching adopted in MadGraph5 to avoid double counting matrix element and parton shower generation of additional jets. 
In the last, while generating backgrounds from the rare SM processes involving weak gauge bosons, we let them decay at the parton level (In this way the helicity information is also retained.). 
The resulting cross sections can be obtained after multiplying the cross sections in Table~\ref{smxsec} by the corresponding branching ratios. Note that only the gauge bosons which decay into $e/\mu$ constitute the backgrounds.


\begin{figure}
 \begin{center}
 \includegraphics[scale=0.37]{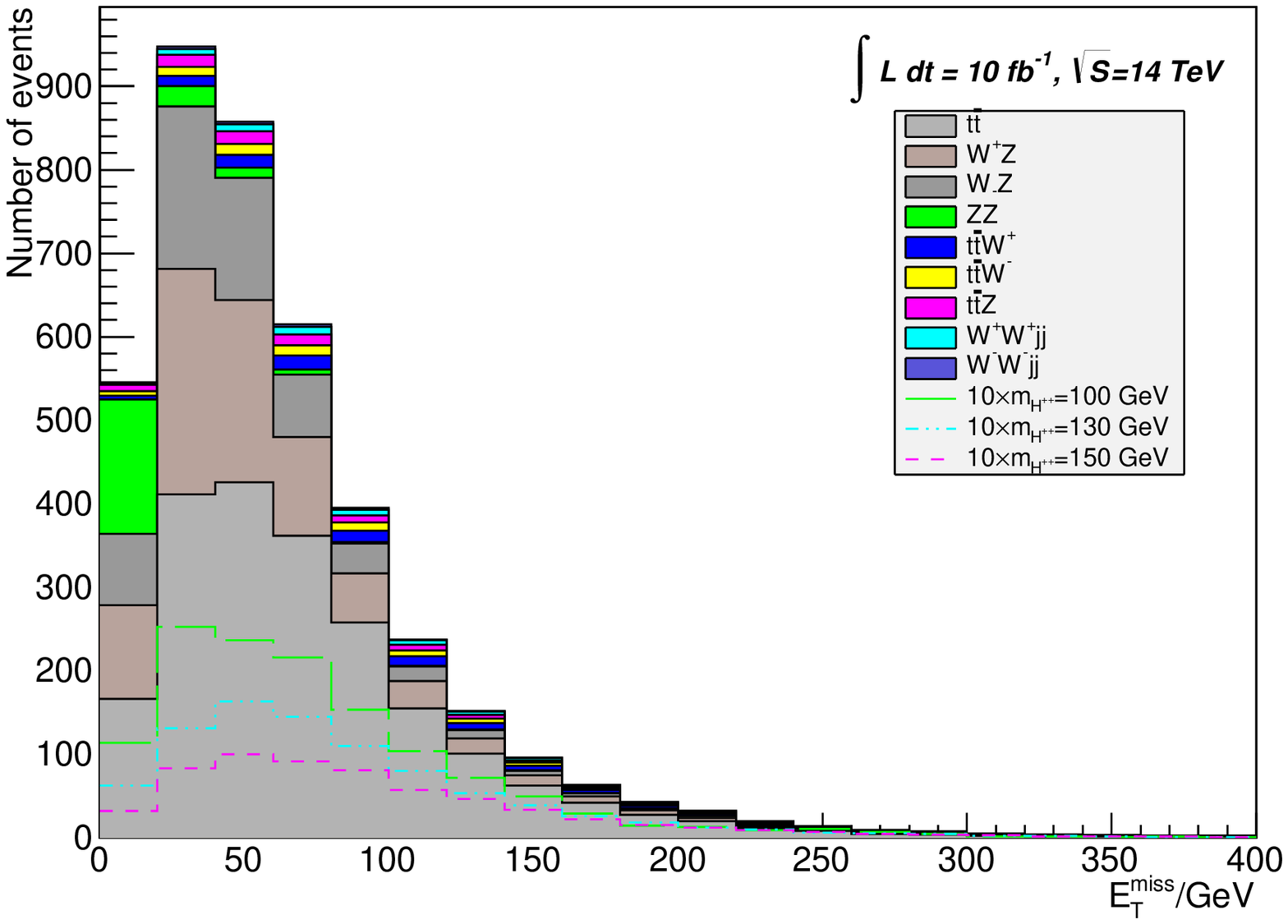}
 \includegraphics[scale=0.37]{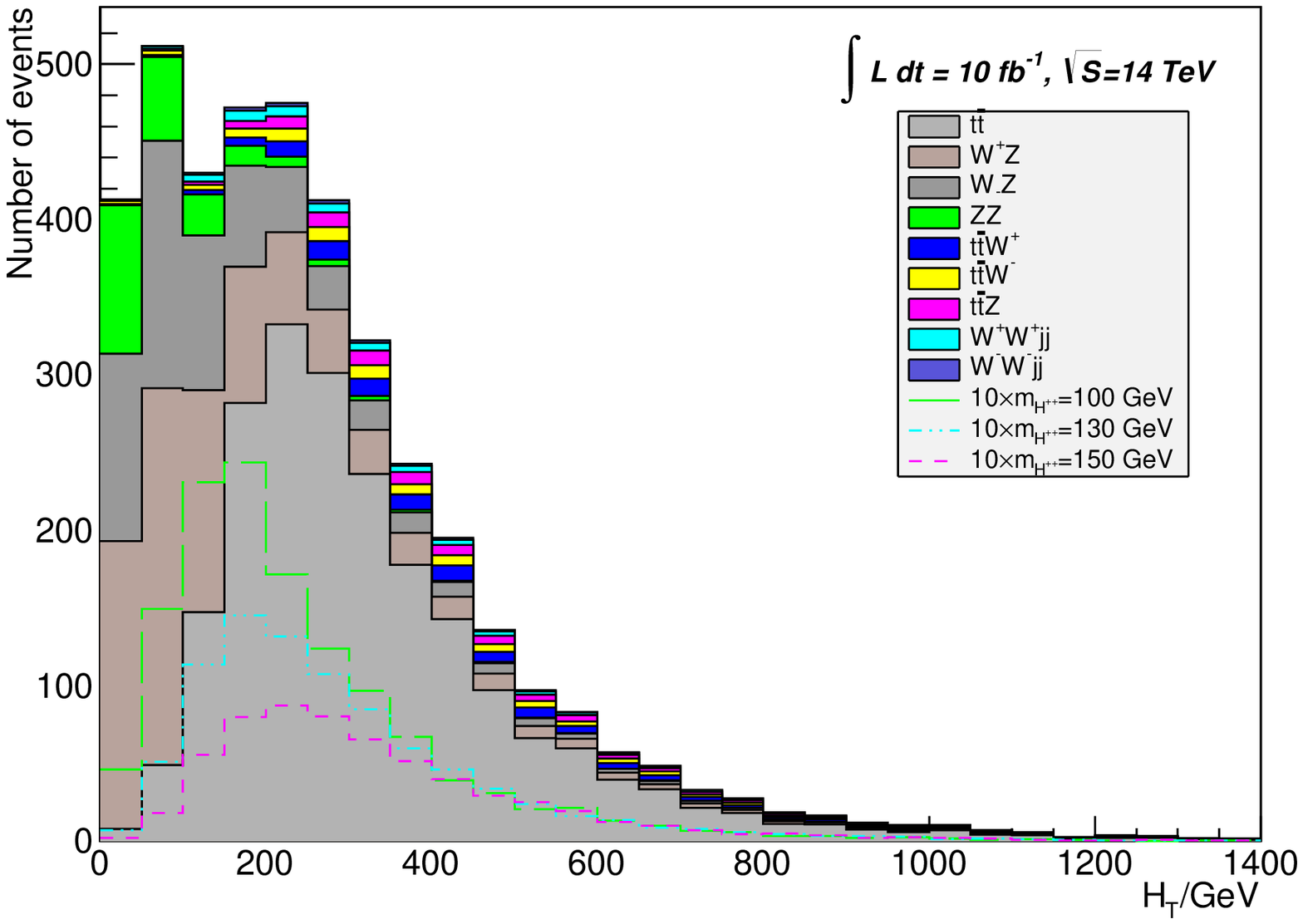}
 \includegraphics[scale=0.37]{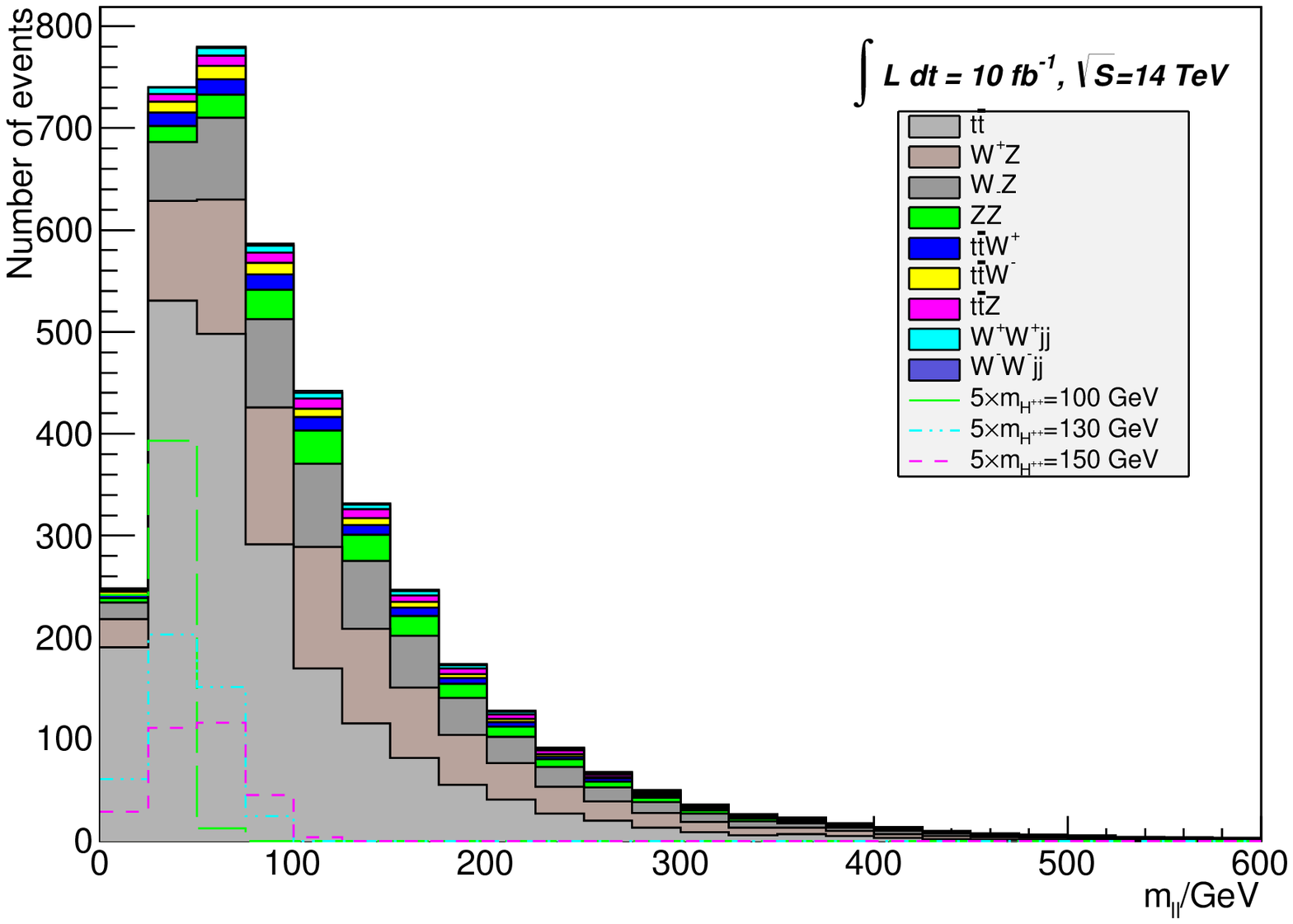}
 \includegraphics[scale=0.37]{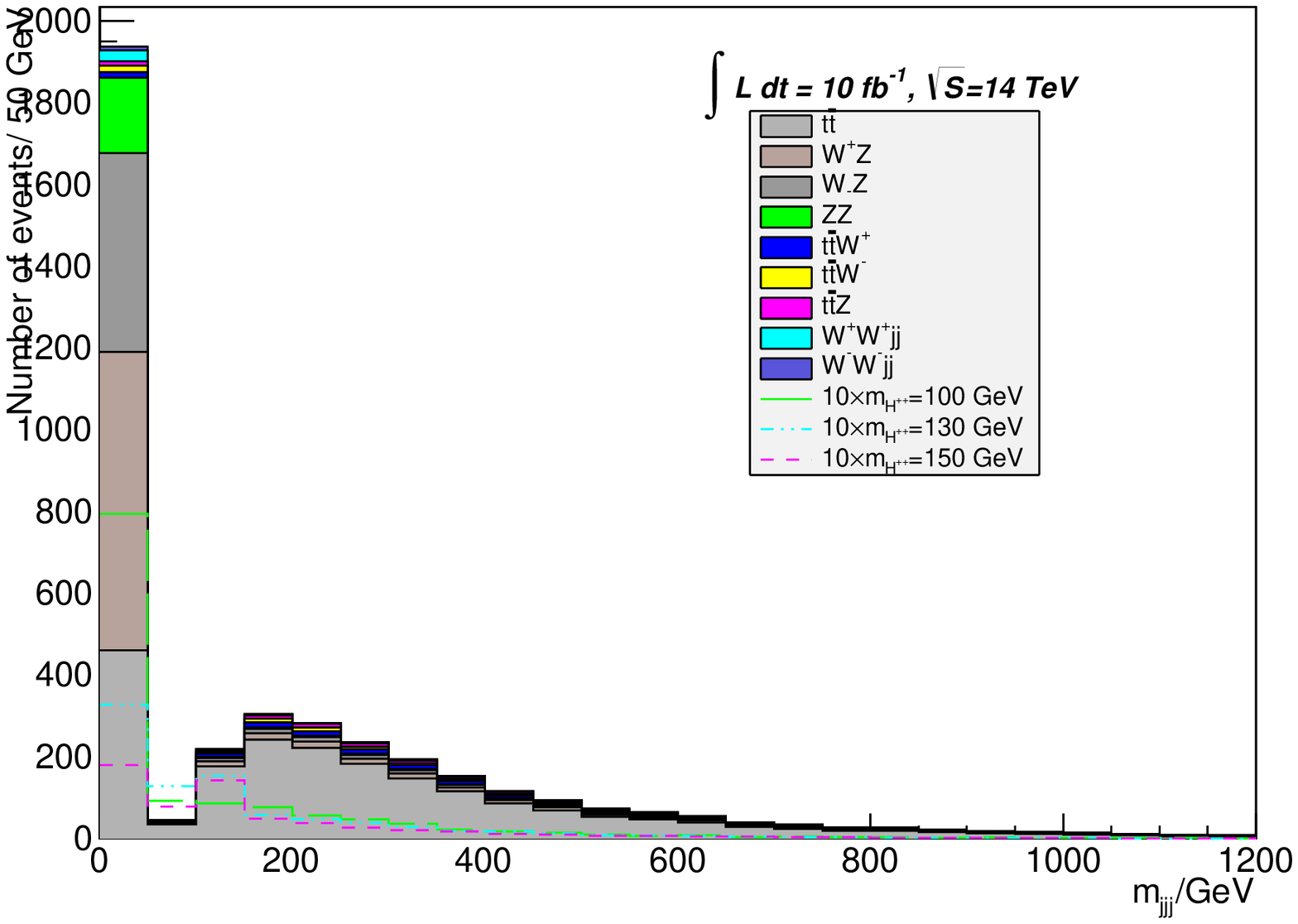}
 \includegraphics[scale=0.37]{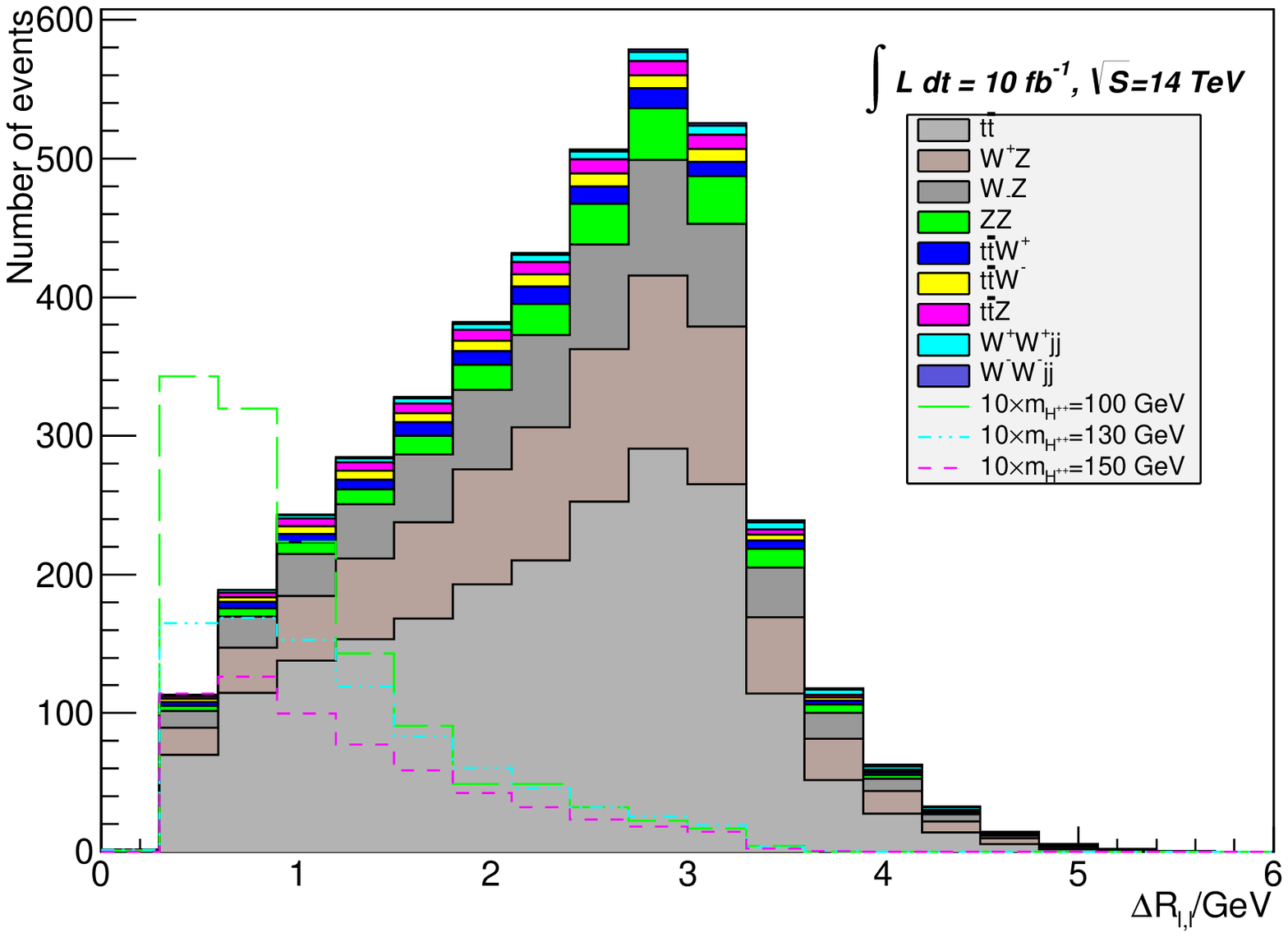}
  \includegraphics[scale=0.37]{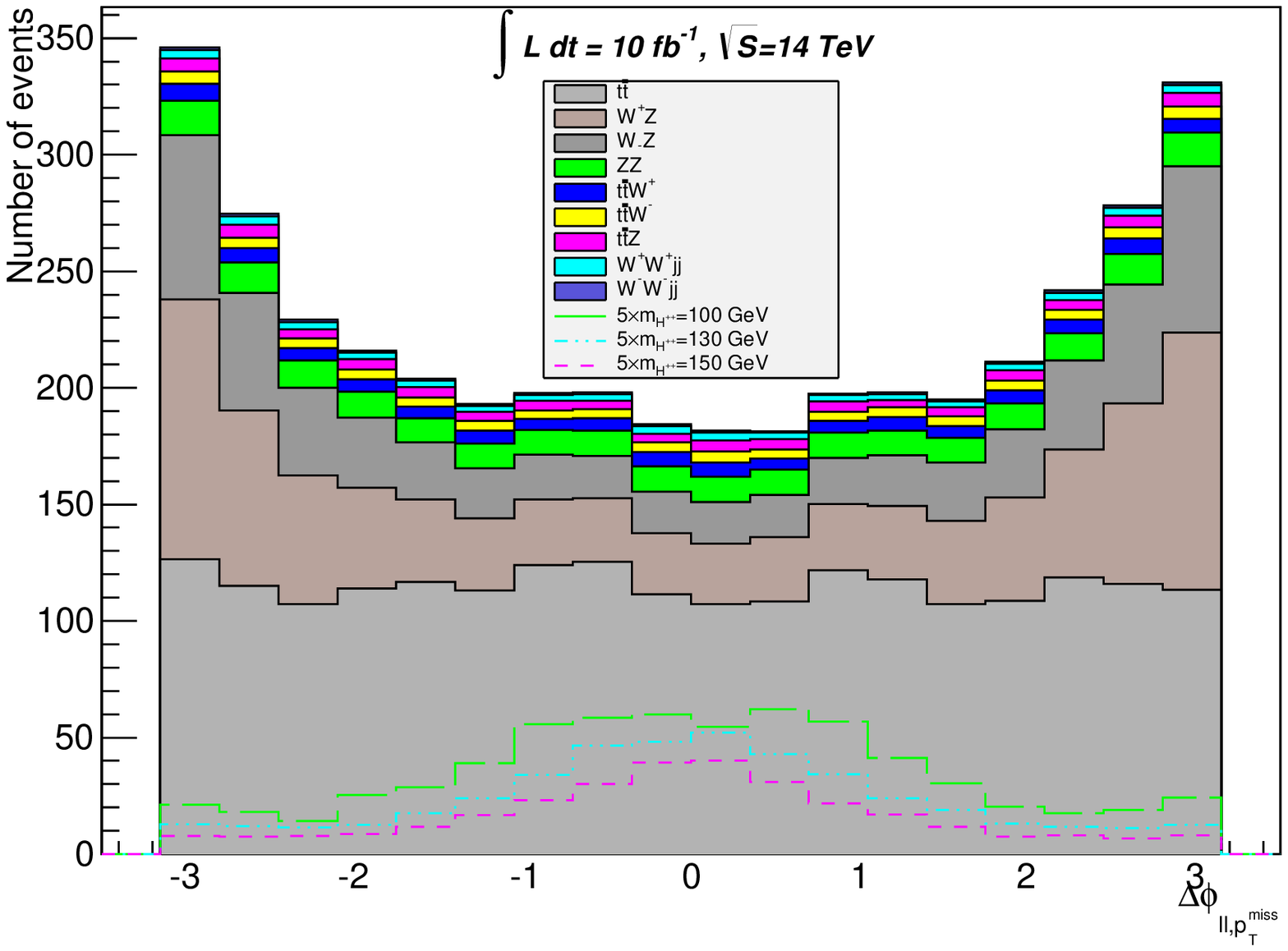}
 \caption{\label{distribution} The distributions for corresponding kinematic variables after SSDL cut of backgrounds and signals. The number of events for signals have been magnified by a ratio as shown in the corresponding figure in order to highlight the distribution of signals.}
 \end{center}
 \end{figure}

With the backgrounds and signal events from simulation, we  consider the event selection procedure in the following:

\begin{itemize}
  \item Events should contain exactly a pair of SSDL and those with additional leptons are vetoed. The leptons are
required to satisfy 
\begin{align}
p_{T,1/2} > 10 ~\text{GeV}, \quad    |\eta| < 2.5.
\end{align}
\item We require at least one jet and moreover no $b$-tagged jets\footnote{In the simulation, we take the $b$-tagging efficiency 0.7~\cite{ATLAS:2012aoa}.} in the signal events. The jets are required to have
       \begin{align}
 p_T>20~ \text{GeV},\quad |\eta|<4.5~.~
\end{align}
\item The LNV decays of $H^{\pm\pm}$ will give small missing energy whereas the hadronic decay of $H^{\pm\pm}$ will give $H_T $ with magnitude proportional to $H^{\pm\pm}$ mass. Thus we require
\begin{align}
E^{miss}_T > 20~ \text{GeV},\quad H_T > 100~\text{GeV}~.~
\end{align}
\item The invariant mass of SSDL pair should be smaller than $H^{\pm\pm}$ mass, i.e.,
\begin{align}\label{ll:cut}
m_{ll}<75 ~\text{GeV}
\end{align}
\item Since $H^{\pm\pm}$ is light, it can be fairly boosted when it is produced at the 14 TeV LHC. As a result, the SSDL pair and the missing transverse momentum will tend to align with each other. Therefore, we impose cuts
\begin{align}
\Delta R(l_1, l_2)  < 1.5,\quad
|\Delta \phi(ll, p^{miss}_T)| < 1.5~,~
\end{align}
where $R(l, l)$ and $\Delta \phi(ll, p^{miss}_T)$ correspond to the angle difference and azimuthal difference between the SSDL system and missing transverse momentum, respectively. 
\item In $H^{\pm\pm}$ decay, two hadronically decaying $W$ bosons produce many jets, especially at the larger $M_{H^{\pm\pm}}$ region. We require that there be at least three jets in the signal events, whose invariant mass should be smaller than 150 GeV.
\end{itemize}

The cuts efficiencies for backgrounds and signals are listed in Table~\ref{background} and Table~\ref{sig}, respectively. 
Since our signal events are generated through the process $p p \rightarrow H^{++}(\to W^+ l \nu,) H^{--}(\to W^{-} j j)$, the events numbers in the 3rd row of Table~\ref{sig} are calculated by $\mathcal{L} \times \sigma(H^{++}H^{--}) \times Br(W \rightarrow \text{hadrons}) \times Br(W \rightarrow l\nu) \times 2 = 2.88 \times  \sigma(H^{++}H^{--})$, where the integrated luminosity $\mathcal{L}=10$ fb$^{-1}$  and the cross section is shown in Fig.1. 
We make some observations from these two tables.
\begin{itemize}
\item As expected, the SSDL cut is the most efficient one to suppress the huge $t\bar{t}$ background, which produces SSDL owing to the heavy flavour quark decay. Even though the requirement of SSDL suppress the $t\bar{t}$ by more than three orders of magnitude, it still stays as the most dominant background for the SSDL signal because of its larger production rate.
\item After SSDL, non-$b$-tagged jet is imposed to further reduce the backgrounds. We also apply the most well studied $E^{miss}_T$ and $H_T$ cuts for comparison, even though they only show very weak discriminative power because of the small $M_{H^{\pm\pm}}$ region. Additionally, it should be noted that the mild cuts of $E^{miss}_T$ and $H_T$ can suppress the non-prompt QCD background where jets can fake as leptons.

\item Since all those signal benchmark points have very small SSDL invariant mass, the signal can be hardly influenced by the cut $m_{ll}< 75$ GeV while all backgrounds turn out to be a few times smaller after this cut.

\item Another feature of the signal process, i.e., alignment of SSDL, can also substantially improve the signal significance. In the background events, SSDL usually comes from two different mother particles decays. So, they tend to have relatively large azimuthal angle difference. In contrast, the lightness of $H^{\pm\pm}$ ensures SSDL and the corresponding transverse missing energy align with each other. This condition can be seen from the corresponding $\Delta \phi(ll, p^{miss}_T)$ distribution shown in the bottom of Fig~\ref{distribution}. 

\item In the last, as seen in the middle right of Fig~\ref{distribution}, the backgrounds either have less than three jets (di-boson background) or have relatively large invariant mass of three leading jets ($t\bar{t}$ background). So after we impose more than 3 jets with invariant mass of three leading jets smaller than 150 GeV ($m_{jjj}< 150$ GeV), all the backgrounds are suppressed by an order of magnitude, while the signals are only a few times smaller. 
\end{itemize}

Increasing $M_{H^{\pm\pm}}$ yields two competitive effects on the cuts. On one hand, the products, both leptons and jets, from a heavier $H^{\pm\pm}$ decay tend to become more energetic, and consequently one has a higher rate of SSDL and a better sensitivity after the $N_j >2$ cut. On the other hand, a larger $M_{H^{\pm\pm}}$ also renders $m_{ll}$ relatively larger, which makes the cut less efficient due to Eq.~(\ref{ll:cut}); moreover, the angular difference cuts also become slightly weaker with larger $m_{H^{++}}$, understood by nothing but less boosted $H^{\pm\pm}$.

To have an impression on the discovery potential,  we calculate the signal significance
\begin{align}
\sigma=S/\sqrt{B+(\beta B)^2}~,~
\end{align}
in which we have assumed { Poisson statistics uncertainty $\sqrt{B}$
and} the systematic error $\beta=5$\%{ \footnote{Because the number of background events in our analysis is very small, the statistical uncertainty is around 35\%. The systematic uncertainty up to $\sim \mathcal{O}(10\%)$ does not affect our results much.} }.
The signal significance for all benchmark points are  given in the last row  of Table~\ref{sig}. From it we are justified to draw such a conclusion: $H^{\pm\pm}$ in the whole region of $100-150$ GeV can be discovered  at the 14 TeV LHC with 10-30 fb$^{-1}$ integrated luminosity.

 We choose the cuts such that our search is most conservative in the whole mass range that we are interested in. As for a specific benchmark point, we can further optimize the corresponding cuts to get a better search sensitivity. For example, for a heavier $H^{\pm\pm}$ one can lower down the $m_{ll}$ cut in Eq.~(\ref{ll:cut}) to get a better signal significance. For $m_{H^{\pm\pm}}=100$ GeV, the $m_{jjj}$ cut can even be dropped; then the signal significance can be as high as 5.3$\sigma$.

\begin{table}[htb]
\begin{center}
\begin{tabular}{|c|c|c|c|c|c|c|c|c|c|} \hline
                         &$t\bar{t}$  & $W_l^{+}Z_l$ & $W_l^{-}Z_l$  & $Z_lZ_l$ & $t\bar{t}W_l^{+}$ & $t\bar{t}W_l^{-}$ & $t\bar{t}Z_l$ & $W_l^{+}W_l^{+}jj$ & $W_l^{-}W_l^{-}jj$  \\ \hline
Events Number & 8433380  & 4278.0 & 2629.7 & 729.9 & 1080.9 & 558 & 733 & 162.1& 70.7 \\ \hline
2SSL & 1978.6  & 499.7 & 314.1 & 56.5 & 88.4 & 52.4 & 35.7& 56.1 & 26.1  \\ \hline
$N_j >0$, $N_b$=0 & 698.4 & 380.3 & 245.4 & 47.9 &14.7 & 8.0 & 5.8 & 53.5 & 24.7 \\ \hline
$E_T^{miss}>20$ &  639.1 & 336.3 & 214.0 & 17.2 & 14.0 & 7.7 & 5.3 & 50.7 & 22.7\\ \hline
$H_T>100$ GeV & 621.7 & 244.0 & 155.6  & 10.5 & 13.9 & 7.6 & 5.3 & 49.5 & 22.1  \\ \hline
$m_{ll}<$75 GeV & 367.3 & 102.2 & 58.6 & 5.5 & 4.5 & 2.3 & 1.7 & 14.2 & 5.1 \\ \hline
$\Delta R(l,l)<1.5$ & 137.2 & 49.3 & 29.2 & 2.9 & 2.2 & 1.4 & 1.1 & 6.2 & 2.7 \\ \hline
$\Delta \phi(ll,p^{miss}_T)<1.5$ & 74.9 & 16.6 & 8.9 & 0.7 & 1.0  & 0.4 & 0.4 & 2.3 & 0.8 \\ \hline
$N_j >2$, $m_{jjj} < 150$ GeV & 6.9 & 0.6 & 0.5 & 0.03 & 0.06 & 0.03 & 0 & 0.05 & 0.02 \\ \hline
\end{tabular}
\end{center}
\caption{\label{background} The cuts flow for backgrounds. The number has normalised to 10 $fb^{-1}$. $W_l$ and $Z_l$ represent the leptonic decays of the gauge bosons. }
\end{table}

\begin{table}[htb]
\begin{center}
\begin{tabular}{|c|c|c|c|c|c|c|} \hline
                         &100 & 110 & 120 & 130 & 140 & 150 \\ \hline
 Ratio required to be excluded & 4.5 & 4.0 & 4.2 & 4.3 & 4.5 & 4.3 \\ \hline
Events Number & 2608 & 1864 & 1365 & 1024 & 786 & 612   \\ \hline
2SSL & 126.3 & 123.3 & 102.9 & 84.5 & 70.2 & 57.8  \\ \hline
$N_j >0$, $N_b$=0 & 114.0 & 112.9 & 94.7 & 78.1 & 64.6 & 53.1 \\ \hline
$E_T^{miss}>20$ & 104.1 & 103.7 & 87.5  & 72.4 & 60.8 & 50.4 \\ \hline
$H_T>100$ GeV & 95.5 & 95.0 & 82.5 & 69.5 & 59.2 & 49.4 \\ \hline
$m_{ll}<$75 GeV & 95.5 & 95.0 & 81.5 & 65.8 & 53.2 & 41.6 \\ \hline
$\Delta R(l,l)<1.5$ & 76.4 & 72.2  & 59.5 & 46.5 & 37.7 & 30.0 \\ \hline
$\Delta \phi(ll,p^{miss}_T)<1.5$ & 61.3 & 56.8 & 46.5 & 36.6 & 29.7 & 23.4  \\ \hline
$N_j >2$, $m_{jjj} < 150$ & 11.2 & 16.3 & 14.4 & 13.6 & 11.3 & 8.8 \\ \hline
$\sigma$ & 3.89 & 5.64 & 4.98 & 4.70 & 3.91 & 3.04 \\ \hline
\end{tabular}
\end{center}
\caption{\label{sig} Cut flow for signal benchmark points. The events number has been normalised to 10 $fb^{-1}$. 
The first row shows
the ratios needed for the production rate so that the benchmark points can be excluded by the CMS search~\cite{Chatrchyan:2013fea}.
In the last row, we show the corresponding signal significances for those benchmark points in our search.}
\label{bksectionbr}
\end{table}

To end up this Section, we comment on possible effects on the $H^{\pm\pm}$ search sensitivity, if we consider different triplet mass spectra. As discussed before, for a non-degenerate spectrum with proper mass splitting, one should include the $H^{\pm\pm} H^\mp$ associated production, which will significantly increase the sensitivity if $H^{\pm\pm}$ becomes the lightest component in the triplet~\cite{H_cas}. In contrast to that, if $H^0$ is the lightest, the cascade decay of $H^{\pm\pm}$ will open; then we can naively expect that the sensitivity will deteriorate due to the decrease of Br$(H^{\pm\pm}\ra WW^*)$~\cite{Chakrabarti:1998qy}.


\section{Conclusion and discussion}\label{conclusion}

The doubly charged Higgs boson $H^{\pm\pm}$ is predicted in a lot of new physics models beyond the SM, and in this paper we implement LHC analysis of $H^{\pm\pm}$ search based on a simplified model with a triplet scalar with hypercharge $\pm1$. The LHC searches for $H^{\pm\pm}$ have been studied widely, but most of the searches focus on the relatively heavy ($\gtrsim 200$ GeV) $H^{\pm\pm}$ dominantly decaying into a pair of SSDL. In this paper we focus on the complimentary region, $m_W\lesssim M_{H^{\pm\pm}}\lesssim 2m_W$. Such light $H^{\pm\pm}$ is hidden at the current colliders as long as the $WW^*$ mode is dominant, which is possible even in the type-II seesaw mechanism when the triplet VEV is significantly large ($\sim$1 GeV) and the mass spectrum of triplet-like Higgs bosons are nearly degenerate. To investigate the LHC prospect of $H^{\pm\pm}$ in that scenario, we performed the detailed signal and background simulations, especially including the non-prompt $t\bar{t}$ background, which is the dominant one but ignored before. We found that $H^{\pm\pm}$ can be discovered at the 14 TeV LHC with 10-30 fb$^{-1}$ integrated luminosity.

\section*{Acknowledgements}
We would like to thank Prof. Eung Jin Chun for helpful discussion.
This research was supported in part by the China Postdoctoral Science Foundation under
grant number 2013M530006 (KZ), by the Natural Science
Foundation of China under grant numbers 10821504, 11075194, 11135003, and 11275246, and by the National
Basic Research Program of China (973 Program) under grant number 2010CB833000 (JL, TL, and YL).

\newpage
\appendix
\section{Decays of the doubly charge Higgs $H^{\pm\pm}$}\label{DHdecay}

In this appendix we present the decay widths of the possible decay modes of $H^{\pm\pm}$. The first is the SSDL mode $H^{\pm\pm}\rightarrow \ell^\pm\ell^\pm$, with decay width
\begin{eqnarray}
  \Gamma(H^{\pm\pm} \rightarrow \ell^{\pm}_i \ell^{\pm}_j) = \frac{|y_{ij}|^2}{4\pi (1 + \delta_{ij})} M_{H^{\pm\pm}}= \frac{|(M_\nu)_{ij}|^2}{8\pi (1 + \delta_{ij})v^2_\Delta}M_{H^{\pm\pm}},
\end{eqnarray}
taking the massless limit of the leptons. Next, for a sufficiently heavy $H^{\pm\pm}$, the di-$W$ mode opens, and the decay width is given by
\begin{eqnarray}
  \Gamma(H^{\pm\pm} \to W^\pm W^\pm)=\frac{g^4v_\Delta^2M_{H^{++}}^3}{64\pi m_W^4}
\left(1-\frac{4m_W^2}{M_{H^{++}}^2}+\frac{12m_W^4}{M_{H^{++}}^4}\right)\beta
\left(\frac{m_W^2}{M_{H^{++}}^2}\right).
\end{eqnarray}
But if $M_{H^{\pm\pm}}<2m_W$, it becomes the three-body mode, i.e., the $WW^*$ mode studied in this paper:
\begin{eqnarray}
 \Gamma(H^{\pm\pm} \to W^\pm W^{\pm *}\to W^\pm f_i \bar{f}_j)&=&\Big(3+ N_C\sum_{q_u,q_d} |V_{q_u,q_d}|^2 \Big)\nonumber\\
 &\times&\frac{g^6M_{H^{++}}}{6144\pi^3}\frac{v_{\Delta}^2}{m_W^2}F
\left(\frac{m_W^2}{M_{H^{++}}^2}\right),
\end{eqnarray}
where the factor $3+ N_C\sum_{q_u,q_d} |V_{q_u,q_d}|^2$ comes from the sum of all possible SM fermions final states. Moreover, $\beta(x)=\sqrt{1-4x}$ and the function $F(x)$ is defined as
\begin{eqnarray}
F(x)&=&47x^2-60x+15-\frac{2}{x}-3(4x^2-6x+1)\log x\nonumber\\
&&+\frac{6(20x^2-8x+1)}{\sqrt{4x-1}}\arccos\left(\frac{3x-1}{2x^{3/2}}\right).
\end{eqnarray}
In the last, when there exists a mass splitting between $H^{\pm\pm}$ and $H^\pm$, the cascade decay mode of $H^{\pm\pm}$ will open; concretely, we works on $\lambda_5<0$ which means $\Delta M= M_{H^{\pm\pm}}-M_{H^{\pm}}>0$. We obtain the two-body decay width for a heavy $H^{\pm\pm}$,
\begin{eqnarray}
  \Gamma(H^{\pm\pm} \to H^\pm W^\pm )=\frac{g^2M_{H^{++}}^3}{16\pi m_W^2}\left[\lambda\left(\frac{m_W^2}{M_{H^{++}}^2},
\frac{M_{H^+}^2}{M_{H^{++}}^2}\right)\right]^{3/2},
\end{eqnarray}
and three-body decay width
\begin{eqnarray}
  \Gamma(H^{\pm\pm} \to H^\pm W^{\pm *}\to H^\pm f_i \bar{f}_j )&=&\Big(3+ N_C\sum_{q_u,q_d} |V_{q_u,q_d}|^2 \Big)\nonumber\\
&\times&\frac{g^4m_{H^{++}}}{256\pi^3} G\left(\frac{M_{H^+}^2}{M_{H^{++}}^2},\frac{m_W^2}{M_{H^{++}}^2}\right),
\end{eqnarray}
for a light $H^{\pm\pm}$. To get the final expression have neglected the mixing between the singly charged Higgs bosons of the triplet and SM double doublet. In the above formulas, the functions $\lambda(x,y)$ and $G(x,y)$ are respectively given by
\begin{eqnarray}
\lambda(x,y)=1+x^2+y^2-2xy-2x-2y,
\end{eqnarray}
and
\begin{eqnarray}
G(x,y)&=&\frac{1}{12y}\Bigg\{2\left(-1+x\right)^3-9\left(-1+x^2\right)y+6\left(-1+x\right)y^2
\nonumber\\
&&-6\left(1+x-y\right)y\sqrt{-\lambda(x,y)}\left[\arctan
\left(\frac{(x-1)\sqrt{-\lambda(x,y)}}{(x-1)^2-y(x+1)}
\right)\right]\nonumber\\
&&-3\left[1+\left(x-y\right)^2-2y\right]y\log x\Bigg\}.
\end{eqnarray}
We have checked that our results are consistent with the ones in Ref. \cite{Yagyu:2012qp}.

\end{document}